\documentstyle[epsfig,prb,aps]{revtex}

\begin{document}
\author{D.J. Singh$^a$, M. Gupta$^b$ and R. Gupta$^c$}
\address{$^a$Center for Computational Materials Science,
Naval Research Laboratory,\\
Washington,
DC 20375-5320 \\
$^b$ Institut des Sciences des Materiaux, Batiment 415, \\
Universite Paris-Sud, 91405 Orsay, France \\
$^c$ Commissariat a l'Energie Atomique, Centre d'Etudes de Saclay, \\
91191 Gif Sur Yvette Cedex, France}
\title{Magnetism and Electronic Structure in ZnFe$_2$O$_4$ and MnFe$_2$O$_4$}
\date{\today}

\twocolumn[\hsize\textwidth\columnwidth\hsize\csname@twocolumnfalse\endcsname

\maketitle

\begin{abstract}
Density functional calculations are used to study magnetic
and electronic properties of the spinel ferrites,
ZnFe$_2$O$_4$ and MnFe$_2$O$_4$. Correct magnetic orderings are obtained.
ZnFe$_2$O$_4$ is predicted to be a small gap insulator in agreement
with experiment.  MnFe$_2$O$_4$ is found to be a low carrier
density half-metal in the fully ordered state. 
However, strong effects on the electronic structure
are found upon partial interchange of Fe and Mn atoms.
This indicates that the insulating character may be
due to Anderson localization associated with the intersite Mn-Fe
disorder.
\end{abstract}


]

Spinel structure ferrites play an
important role in technologies because
a large subset of these materials are
room temperature, insulating ferromagnets.
While the basic physics underlying the magnetic orderings has be
understood since the work of Goodenough and co-workers,
\cite{good-kan1,good-kan2} a fundamental predictive understanding of
issues like gap formation is not yet well established.
Here we report density functional calculations of the electronic
structure of ZnFe$_2$O$_4$ and MnFe$_2$O$_4$, focussing mainly on the latter.
MnFe$_2$O$_4$
differs from ZnFe$_2$O$_4$ in an important aspect.
As prepared, it invariably has significant disorder between
the Mn and Fe sites. \cite{hastings}
This intersite disorder amounts to approximately
20\% Fe on the Mn site in normally prepared
stoichiometric samples.
This disorder has caused some
confusion both in the analysis of the dependence of magnetic
properties on sample size
\cite{kulkarni,zaag2,chen,flores}
and in understanding the temperature
dependence of transport quantities.
\cite{simsova1,simsova2}
MnFe$_2$O$_4$ is insulating.
Small gap values from 0.04 -- 0.06 eV (Ref. \onlinecite{flores})
have been extracted from transport data determined from eddy currents.

In the generalized gradient approximation (GGA),
\cite{perdew}
a small insulating gap,
consistent with experiment, is found in ZnFe$_2$O$_4$, while ideal
stoichiometric MnFe$_2$O$_4$ predicted to be a low carrier density
half-metal. However, total energy calculations show that
partial inversion is strongly favored in MnFe$_2$O$_4$ so the ideal
ordered phase cannot be made, at least not by standard methods.
The calculations show a strong enough coupling
of the electronic states near the Fermi level $E_F$ to intersite cation
disorder in MnFe$_2$O$_4$ to exceed the criterion for Anderson localization
in this material. We propose that this may underlie the
insulating behavior of MnFe$_2$O$_4$, and note that it is consistent
with a number of experimental observations, like the difficulty
in determining gap values spectroscopically, and the variation
in the transport gap with temperature and sample preparation.

Our results were obtained using well converged linearized
augmented planewave (LAPW) calculations. \cite{singh-book} Details are
given in our prior paper, where some results for ZnFe$_2$O$_4$ are reported.
\cite{singh-zn}
There, we found that the electronic structure is highly sensitive to the
structural parameters used, particularly, the internal O coordinate $u$ of
the spinel structure. For ZnFe$_2$O$_4$ we used the experimental value
determined from structural refinements. MnFe$_2$O$_4$ has not been made
in non-inverted form, so we instead used total energy minimization.
In this way we obtained $u$=0.381 and $a$=8.49\AA.
With these and no inversion, we find the ferrimagnetic state
to be the ground state. Specifically, we did
total energy calculations for the ferrimagnetic state, a ferromagnetic state
and antiferromagnetic $A$ and $B$ sublattices.
The ferrimagnetic configuration
was lowest in energy, {\it e.g.} by 2.03 eV per cell (2 Mn and 4 Fe ions)
relative to the ferromagnetic. Details will be given in a longer report.
\cite{singh-mn}

\begin{figure}[tbp]
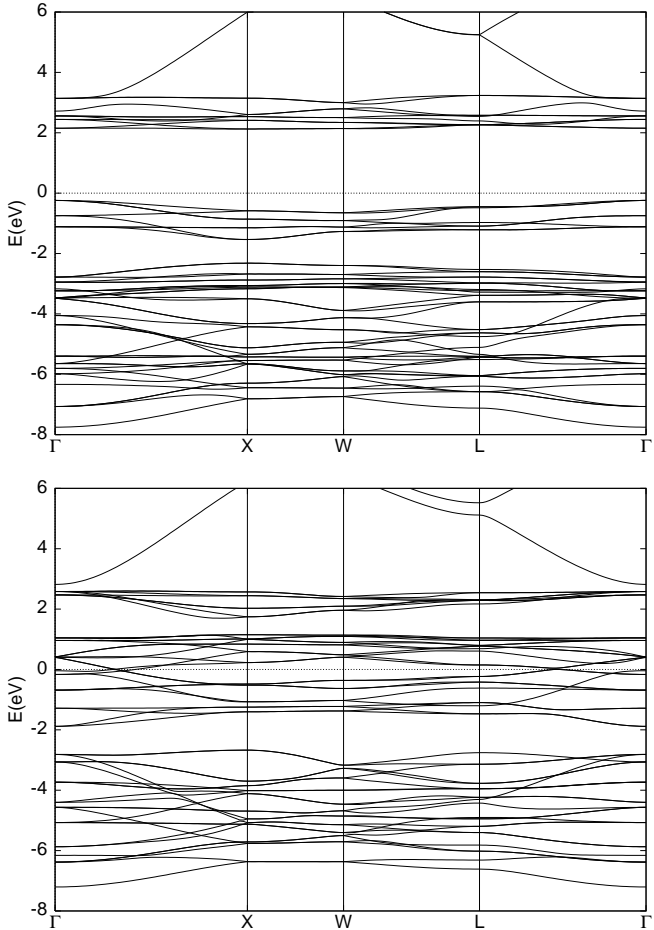

\centerline{\epsfig{file=band2560-up.epsi,angle=270,width=0.99\linewidth}}
\vspace{0.125in}
\centerline{\epsfig{file=band2560-dn.epsi,angle=270,width=0.99\linewidth}}
\vspace{0.125in}
\setlength{\columnwidth}{3.2in} \nopagebreak
\caption{
Band structure of ferrimagnetic MnFe$_2$O$_4$.
$E_F$ is at 0.
Note the half-metallic gap at $E_F$ in the majority
(top panel) but not in the minority (bottom panel).
}
\label{bands-ground}
\end{figure}

\begin{figure}[tbp]
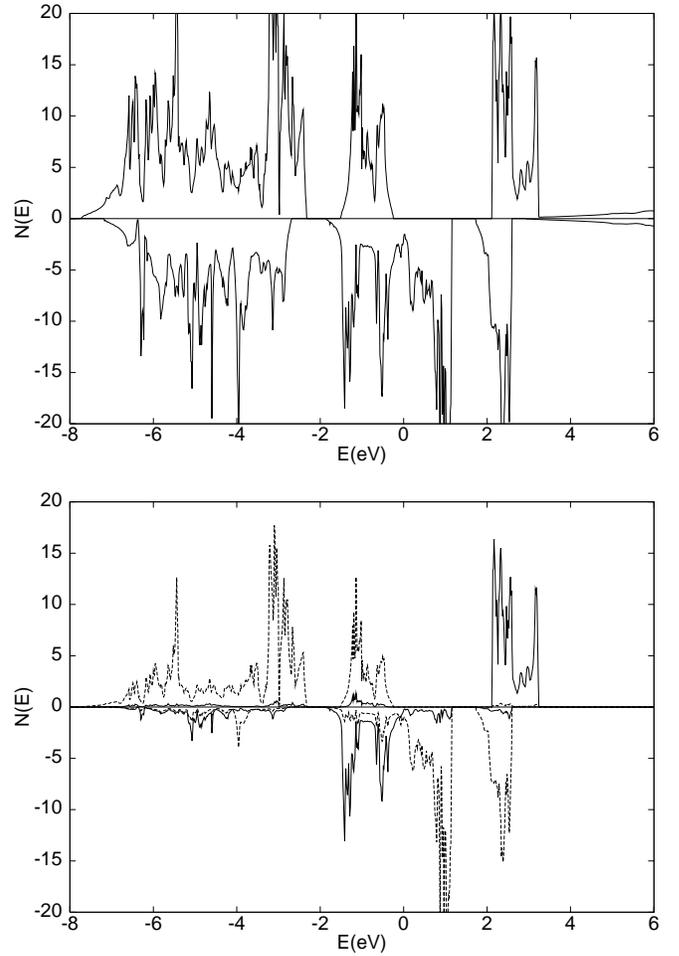

\centerline{\epsfig{file=dos2560.epsi,angle=270,width=0.99\linewidth}}
\vspace{0.125in}
\centerline{\epsfig{file=dos2560-mn-fe.epsi,angle=270,width=0.99\linewidth}}
\vspace{0.125in}
\setlength{\columnwidth}{3.2in} \nopagebreak
\caption{
Electronic DOS corresponding to Fig. \ref{bands-ground} (top),
and projections onto the Mn (solid) and
Fe (dashed) LAPW spheres (bottom).
The majority spin is shown
above the axis and the minority below.
}
\label{dos-ground}
\end{figure}

\begin{figure}[tbp]
\centerline{\epsfig{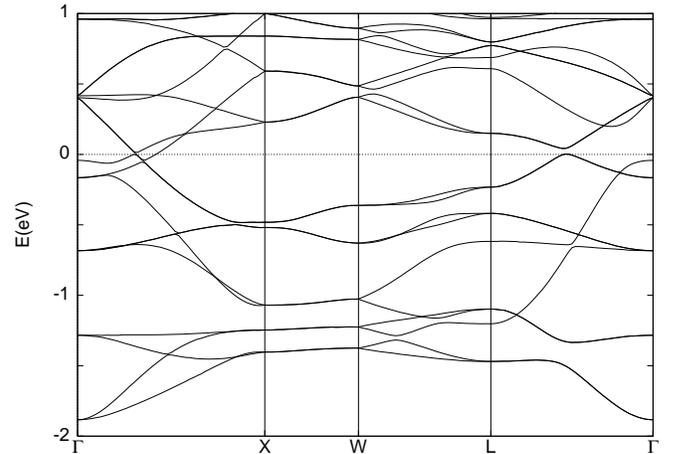}}
\vspace{0.125in}
\setlength{\columnwidth}{3.2in} \nopagebreak
\caption{
Blowup around $E_F$ of the
minority spin band structure as in Fig. \ref{bands-ground}.
}
\label{minority-blowup}
\end{figure}

In an ionic model, Fe has a trivalent high spin $d^5$ configuration,
while both Zn and Mn are divalent, so Mn is also $d^5$.
We find a spin magnetization
of exactly 10 $\mu_B$ per unit cell in the ferrimagnetic state, as in
the ionic model. As seen in the band structure (Fig. \ref{bands-ground}),
this is due to a half-metallic state. The density of states (DOS) and
projections (Fig. \ref{dos-ground}) show that the ionic model is
followed,
though there is some reduction in moments due to hybridization.
There is a
majority spin gap but only a pseudogap in the
minority channel. The majority channel is described by the
ionic model. It has O $2p$ bands from
$\sim$ -8 eV (relative to the Fermi energy, $E_F$)
to $\sim$ -3 eV. Fe $t_{2g}$ bands overlap
the top of these and extend to -2.3 eV. These are separated by a
clean crystal field gap from an Fe $e_g$ manifold, which goes
from -1.5 to -0.2 eV. The unoccupied Mn $d$ bands are above +2 eV
and show a smaller tetrahedral crystal field with
overlapping $e_g$ and $t_{2g}$ manifolds.
As mentioned, the minority spin channel is metallic. This is a result
of a manifold of 22 bands lying between -2 eV and 1 eV
(Fig. \ref{minority-blowup}).
The lower part of this manifold
(from -2 to -1 eV) is from 4 Mn $e_g$ bands, which are
below the $t_{2g}$ bands in the Mn tetrahedral crystal field. The
next set (of 6), extending to $E_F$, is Mn $t_{2g}$ derived, while the
remaining 12 bands are from the Fe $t_{2g}$.
There is a pseudogap at $E_F$. This
separates the Mn $t_{2g}$ and Fe $t_{2g}$ states.
The states at
$E_F$ come from the overlap of a two fold degenerate band that
disperses downwards from 0.4 eV at $\Gamma$ with two singly degenerate
bands starting just below $E_F$ at $\Gamma$ and dispersing upwards.
The upward dispersing bands (from $\Gamma$) come from
Fe $d$ states, while the downward dispersing band is Mn derived.
The downwards (hole) and upwards (electron) bands cross very close
to $E_F$ on the $\Gamma$-X line. Their anti-crossing is
sufficiently strong that a true gap opens up between them along
the other symmetry lines (see {\it e.g.} the $\Gamma$-L line. The
result is a semimetallic half-metal, with a large band
gap over 2 eV in the majority channel and small electron and hole
Fermi surfaces along the $\Gamma$-X direction
in the minority channel. 

The calculated band structure needs to be reconciled with the
experimentally observed insulating nature of MnFe$_2$O$_4$.
For this we focus on the region near $E_F$, which is relevant to transport.
We calculated the band structure with the experimental crystal structure, 
but still having no inversion.
This amounts to an
oxygen shift of 0.06\AA~and a very slight lattice expansion.
With this shift, the band topology changes
because of a 0.3 eV upwards shift of
the (previously) first occupied band from $E_F$ in the calculated structure.
This Fe derived
band then becomes unoccupied at $\Gamma$ and the smaller electron Fermi
surface is removed while a new hole surface is introduced due to the strong
downwards dispersion of this band along $\Gamma$-L.
The value of $u$ for the experimental structure is
an average diffraction value for a disordered crystal. Presumably,
the main contribution to the shift in $u$ is the difference in ionic
radii of Mn and Fe, which is a local effect. As such, it is reasonable
to suppose that the local static variation in the oxygen positions in the
disordered material will be at least as large. With this,
band shifts larger than the effective
$E_F$ of the hole and electron pockets should be present. This is the
criterion for Anderson localization. Furthermore, in addition to
band shifts due to O relaxation around intersite defects, there will be
a contribution due to the different ionic potential of the anti-site
species. Here this is the stronger effect.
To qualitatively characterize the changes in ionic
potential, we did calculations with the MnFe$_2$O$_4$ structure, but with
one or two Fe atoms interchanged with Mn in the unit cell (containing
4 Fe and 2 Mn atoms). These interchanges strongly modify the electronic
structure near $E_F$. For example, interchanging a single Fe-Mn pair in
the unit cell destroys the half-metallic electronic structure by
introducing a narrow majority spin peak derived mostly from the $d$ electrons
of the $B$-site Mn atom at $E_F$. Meanwhile, the minority spin bands
are shifted so that the pseudogap at $E_F$ becomes a true gap, though
because of the majority occupation it is shifted to slightly (0.2 eV) below
$E_F$. With both Mn on the $B$-sublattice, this minority gap is further
opened up to 0.6 eV.
The shifts in band edge electronic states are
clearly well above the threshold for Anderson localization.

Here, we froze the atomic positions
at the structure for non-inverted MnFe$_2$O$_4$, thus
neglecting relaxation around the anti-site defect. Even so,
the calculated total energies for the partially inverted cells were
lower than for the ideal structure; we get an
energy lowering of 8 mRy per Mn for a single interchange and 12 mRy
per Mn for a double interchange. While the neglected
relaxation is no doubt important
it can only raise these energy differences (further disfavoring the
non-inverted state).
Presumably, a large ingredient in the relaxation will be a competition
between neighboring metal atoms for metal-oxygen bond lengths.
This interaction will be repulsive between neighboring antisites,
and if strong enough may limit the amount of inversion.
In any case,
the calculated energetics mean
that at least partial inversion is energetically
favored in MnFe$_2$O$_4$ with a rather high energy scale (
at least 8 mRy per pair {\it i.e.}
$>$ 1250K), so we conclude that it is not possible to make
non-inverted MnFe$_2$O$_4$ by equilibrium methods at accessible
temperatures. This is unfortunate since our results show
that if it could be made and the insulating gap is really due to
Anderson localization, then non-inverted MnFe$_2$O$_4$ would be a
low carrier density half-metal with a Curie temperature well
above room temperature, and therefore of interest for
spintronic applications. One may speculate that epitaxial growth, with
strain, may favor different inversion levels. Perhaps under suitable
conditions this can be exploited to grow films that are not inverted
at least near an interface -- a possibility that deserves
experimental study.

It seems plausible that this Anderson localization
mechanism for insulating behavior may be operative in MnFe$_2$O$_4$. If so,
localized states should exist in the gap region around $E_F$. These
may be observable thermodynamically and spectroscopically.
Also, the effective value of the
insulating gap will be sensitive to the strength of the disorder.
Of course, it is possible that the insulating nature of MnFe$_2$O$_4$
is due to Hubbard correlations.
Though the behavior of
the kind of multi-site, multi-band Hubbard type model that would be needed
to describe MnFe$_2$O$_4$ is no doubt complex, band shifts controlled
by $U$ or $U-\Delta$, where $\Delta$ is the onsite LSDA Hund's exchange
splitting, may be expected. It seems that in this case gaps too large
compared to experiment may be expected.
It will be of interest to
measure thermodynamic and spectroscopic properties of MnFe$_2$O$_4$
with various levels of inversion to shed light on this issue.

This work is supported by ONR. Computations were done at the ASC
and IDRIS computer centers.

\end{document}